\documentclass[pre,10pt,twocolumn,showpacs]{revtex4}
\usepackage{amssymb}
\usepackage{graphicx}

\begin{document}

\title{Morphological transition between diffusion-limited and ballistic aggregation growth patterns}

\author{S. C. Ferreira Jr.} \email{silviojr@ufv.br}
\affiliation{Departamento de F\'{\i}sica, Universidade Federal Vi\c{c}osa,
36571-000, Vi\c{c}osa, MG, Brazil}
\author{S. G. Alves}
\author{A. Faissal Brito}
\author{J. G. Moreira}
\affiliation{Departamento de F\'{\i}sica, Universidade Federal de Minas
Gerais, CP 702, 30161-970, Belo Horizonte, MG, Brazil}


\begin{abstract}
In this work, the transition between diffusion-limited and
ballistic aggregation models was revisited using a model in
which biased random walks simulate the particle
trajectories. The bias is controlled by a parameter
$\lambda$, which assumes the value $\lambda=0$ ($1$) for
ballistic (diffusion-limited) aggregation model. Patterns
growing from a single seed were considered. In order to
simulate large clusters, a new efficient algorithm was
developed. For $\lambda \ne 0$, the patterns are fractal on
the small length scales, but homogeneous on the large ones.
We evaluated the mean density of particles $\overline{\rho}$ in the region
defined by a circle of radius $r$ centered at the initial
seed. As a function of $r$, $\overline{\rho}$ reaches the asymptotic
value $\rho_0(\lambda)$ following a power law
$\overline{\rho}=\rho_0+Ar^{-\gamma}$  with a universal exponent
$\gamma=0.46(2)$, independent of $\lambda$. The asymptotic value has
the behavior $\rho_0\sim|1-\lambda|^\beta$, where $\beta= 0.26(1)$.
The characteristic crossover length that
determines the transition from DLA- to BA-like scaling
regimes is given by $\xi\sim|1-\lambda|^{-\nu}$, where
$\nu=0.61(1)$, while the cluster mass at the crossover follows a power law $M_\xi\sim|1-\lambda|^{-\alpha}$, where $\alpha=0.97(2)$. We deduce the scaling relations $\beta=\nu\gamma$ and $\beta=2\nu-\alpha$ between these exponents.
\end{abstract}

\pacs{61.43.Hv,05.40.Fb,47.53.+n,47.54.+r}

\maketitle

\section{Introduction}
\label{intro}

The pattern formation in nonequilibrium processes is a longstanding problem
in Statistical Physics \cite{Meakinbook,Vicsekbook,BarabasiBook}. In special,
the diffusion-limited aggregation model (DLA) \cite{Witten} is a noteworthy
example in which a very simple algorithm generates disorderly fractal
clusters. This model was related to several physical and biological
applications, such as electrodeposition \cite{Matsushita}, viscous fingering
\cite{Maloy}, bacterial colonies \cite{Matsushita2}, and neurite formation
\cite{Caserta}. In DLA model, particles released at a point distant from the
cluster execute random walks until they find a neighbor site of the cluster
and irreversible stick in this site. If the random walks are replaced by
ballistic trajectories at random directions, we have the ballistic aggregation
(BA) model \cite{Vold}. In contrast to DLA, the BA model generates disordered
nonfractal clusters with nontrivial scaling properties
\cite{Vicsekbook,Liang}.

Due to its importance as a fundamental model, several generalizations of DLA
model were proposed \cite{Meakinbook,Vicsekbook}. In particular, those models
in which the particle trajectories are biased random walks were investigated
\cite{MeakinPRB, KimPRA, KimPRE, NagataniPRA, CastroPRE,Kim}. In these models,
on the short length scales the particle trajectories are common random walks
with fractal dimension 2.0, whereas on the longer length scales the bias
becomes dominant and the dimensionality of the walk is 1.0. Clusters grown
using this type of walk must behave like DLA model on the short length scales,
while nonfractal patterns are observed on the longer ones. Consequently, the
mass of a cluster of size $l$ is given by
\begin{equation}
    M(l)= l^{d_f}f(l/\xi),
\label{eqM}
\end{equation}
in which
\begin{equation}
f(x)\sim\left\{ \begin{array}{l}
                const. \mbox{~if~} x\ll 1\\
                x^{d-d_f} \mbox{~if~} x\gg 1.
                \end{array}\right.
\label{eqf}
\end{equation}
Here, $d$ is the space dimension, $d_f$ is the DLA fractal dimension, and
$\xi$ is the crossover radius from DLA- to BA-like scaling regimes. This idea
was firstly considered by Meakin \cite{MeakinPRB}. In his model, the
simulations start with a single seed at the center of a square lattice and the
drift of all trajectories is in a fixed lattice direction. Along the walk, the
particle is moved one lattice unity in the drift direction with probability
$P$, or moves to one of its four next-neighbor sites with probability $1-P$.
The model generates patterns with a growth tendency in the opposite direction
of the drift. The author argues that the crossover from the DLA-like structure
on the short length scales to a compact structure on the longer ones is
characterized by a length $\xi\sim P^{-1}$. However, using a renormalization
group approach, Nagatani found $\xi\sim P^{-1/(d-d_f)}$ \cite{NagataniPRA88}.
Kim \textit{et. al.} \cite{KimPRA,KimPRE} studied lattice models with a global
drift to the seed, in which the particles have a higher probability to move to
the nearest neighbor representing the shortest distance away from the seed.
The pattern morphologies are ruled by the lattice anisotropy and their fractal
dimension is 1. Nagatani \cite{NagataniPRA} considered the effects of positive
and negative radial drifts in DLA model. In the positive case the cluster
fractal dimension is asymptotically 2, while eccentric patterns with dimension
1 were found for the negative case. Other models \cite{CastroPRE,Kim} consist
of the deposition processes on a $d$-dimensional substrate in which the walk
drift is the substrate direction.

In the present paper, we are interested in the transition from DLA to BA
models when the random or ballistic trajectories of DLA and BA models,
respectively, are replaced by biased random walks with a random drift
direction. The central concern of this work is the fact that all real fractals
exhibit scaling only on limited ranges and, consequently, a quantitative
analysis of both experiments and simulations demands a deep understanding of
these crossovers. The outline of the paper is the following. In Sec.
\ref{models}, the model and the respective computational algorithm are
described. In Sec. \ref{results}, the simulational results are presented and
discussed. Finally, some conclusions are drawn in Sec. \ref{conclusions}.

\section{Models and Methods}
\label{models}

As in the original DLA and BA models, at the beginning of the simulations a
unique seed localized in the center of the lattice constitutes the cluster.
Then, particles are sequentially released at a circle distant from the cluster
and execute biased random walks. The distance between the center of the
lattice and the launching circle is denoted by $R_l$. The biased walks are
defined by
\begin{eqnarray}
x_{n+1}=x_n+\cos(\varphi + \lambda \theta_n) \nonumber \\
y_{n+1}=y_n+\sin(\varphi + \lambda \theta_n),
\label{bias_walk}
\end{eqnarray}
where $x_n$ and $y_n$ are the particle coordinates at the n$th$ step of the
walk, $\varphi$ is a random angle that defines the bias direction, $\lambda
\in [0,1]$  is the parameter that controls the random component of the
trajectories, and $\theta_n$ is a random direction. The variables $\varphi$ and
$\theta_n$ are in the range $[-\pi,\pi]$. Notice that
$\varphi$ is defined at the beginning of the walks, whereas $\theta_n$ assumes
random values for each walk step. One can see that the particular cases
$\lambda=0$ and $\lambda=1$ recover the BA and DLA models, respectively. If
the particle visits a site neighboring the cluster it irreversibly joins to
this site. However, if the distance between the particle and the cluster is
too large, i. e., larger than a killing radius $R_k$, the particle is excluded
and a new one is released at the launching circle. In order to determine when
a walker is neighboring a cluster site, its lattice position was defined as
the next integer value of its real coordinates defined by Eq.
(\ref{bias_walk}). The majority of the results presented in this work refer to
the square lattice version of the model.

\begin{figure}[hbt]
\begin{center}
\resizebox{8.0cm}{!}{\includegraphics{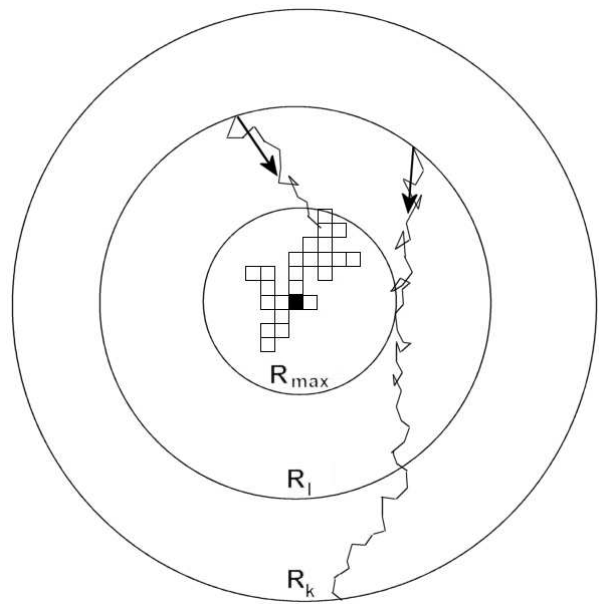}}
\end{center}
\vspace{-0.2cm}
\caption{Schematic representation of the model.}
\label{modeldba}
\end{figure}

The previously introduced variables ($R_l$ and $R_k$) must be as large as
possible. However, computational limitations restrict the use of $R_l$ and
$R_k$ values. We defined $R_l=R_{max}+R_0$, where $R_{max}$ is the maximum
distance from the center of the lattice of a particle belonging to the
cluster. For DLA and BA models $R_0$ can be of the order of some lattice units
\cite{Meakinbook, Vicsekbook}. However, for the model with biased random walks
the patterns morphologies are strongly dependent on this value. Our tests
suggest that the patterns become insensitive to $R_0$ variations when
$R_0>300$, in agreement with the values adopted by Kim \cite{Kim} for a model
of deposition of biased random walks on a substrate. Thus, $R_0=400$ were used
in all simulations. The killing radius $R_k$ must be $10$ to $100$ times
$R_{max}$ for very large DLA clusters \cite{Meakinbook} whereas a $R_k$ only
some lattice units larger than $R_{max}$ is necessary for the BA model. Due to
the bias present in the random walks, we used the same strategy adopted by Kim
\cite{Kim}, i. e., $R_k=2R_{max}+R_0$. Fig. \ref{modeldba} illustrates two
tentatives, one successful and the other frustrated, to add a new particle to the
cluster.

\begin{figure}[hbt]
\begin{center}
\resizebox{8.5cm}{!}{\includegraphics{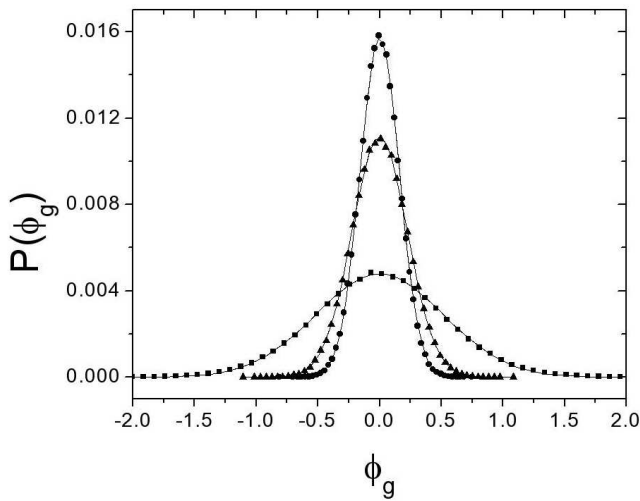}}
\end{center}
\vspace{-0.7cm}
 \caption{Angle distributions for biased random walkers with
preferential direction $\varphi=0$ and the correspondent Gaussian fits. The
curves correspond to $\lambda=0.99$ and $R_s=200$ (squares), $\lambda=0.99$
and $R_s=1000$ (triangles), and $\lambda=0.90$ and $R_s=200$ (circles). Other
parameter sets used in the simulations provide fits better than those shown in
this figure. For all couples $\lambda$ and $R_s$ used in the simulations a
correlation coefficient $r^2 \ge 0.999$ was required.} \label{gaussfit}
\end{figure}

To analyze the transition between BA and DLA it is necessary to simulate large
clusters using lattices containing up to $10^4\times 10^4$ sites, specially
when $\lambda \lesssim 1$. Consequently, the computational time becomes
prohibitive and an efficient algorithm is necessary. A technique commonly used
to simulate large DLA and related models is to allow the particles execute
long steps at random directions if they are far from the cluster \cite{Ball,
Meakin, Tolman, Ferreira}. This procedure is correct because the probability
of a random walker crosses the circle centered on its initial position in a
given angle $\phi$ is uniformly distributed in the interval $[-\pi,\pi]$.
However, for the biased random walkers this is not true. Indeed, the
probability density distributions are concentrated around the direction
$\varphi$. For $\lambda$ not very close to the unity or for large steps, the
probability distributions are very well fitted by Gaussian curves centered at
$\varphi$, as illustrated in Fig. \ref{gaussfit}. Using this fact, the
following procedure was adopted. If the distance between the random walker and
the cluster is larger than a value $R_s+\delta$, it executes a jump of length
$R_s$. Thus, a long jump can not lead the walker to a distance smaller than
$\delta$. The jump direction is $\phi=\varphi+\phi_g$, where $\phi_g$ is a
random number between $-\pi$ and $\pi$ selected from a Gaussian distribution
\begin{equation}
P(\phi_g)= \frac{1}{\sigma\sqrt{\pi/2}} \exp\left(-2\frac{\phi_g^2}{\sigma^2}\right).
\end{equation}
In order to obtain the Gaussian width $\sigma$ for each couple $\lambda$ and
$R_s$, a large number of biased random walks ($10^6-10^7$) were simulated and
a histogram of the probabilities built (Fig. \ref{gaussfit}). Then, the
$\sigma$ value can be determined using least square Gaussian fits. The quality
of fits is improved as larger $R_s$ values and smaller $\lambda$ values are
used. Therefore, the $R_s$ values should be sufficiently large for reproduce
good fits, in special for $\lambda\lesssim 1$. Furthermore, several $R_s$
values can be used in a same simulation improving the algorithm efficacy. We
used two values: $R_s=200$ and $R_s=1000$. The $\sigma$ values used in the
simulations are shown in table \ref{sigmas}. Also, $\delta=20$ were used. All
tests show that the growth patterns are not sensible to the $\delta$ value.
It is worth to note that the Gaussian distribution is not normalized in the
interval $[-\pi,\pi]$ and, obviously, this is not the actual angle
distribution for the present problem. However, the very good fits to the
angle distributions justify the use of the Gaussian functions.

\begin{table}[hbt]
\begin{tabular}{ccc}
\hline
\hline
$\lambda$  &  $~~\sigma (R_s=200)~~$   & $~~\sigma (R_s=1000)~~$\\ \hline
0.100      &  0.02585              & 0.01202 \\
0.300      &  0.07591              & 0.03413 \\
0.500      &  0.12535              & 0.05613 \\
0.700      &  0.18206              & 0.08139 \\
0.900      &  0.31933              & 0.14216 \\
0.950      &  0.45109              & 0.20057 \\
0.990      &  1.03575              & 0.45130 \\
0.995      &  1.53278              & 0.63451 \\
\hline \hline
\end{tabular}
\caption{$\sigma$ values determined with the Gaussian least square fits.}
\label{sigmas}
\end{table}

\section{Results and Discussions}
\label{results}

\begin{figure}[hbt]
\begin{center}
\resizebox{8.5cm}{!}{\includegraphics{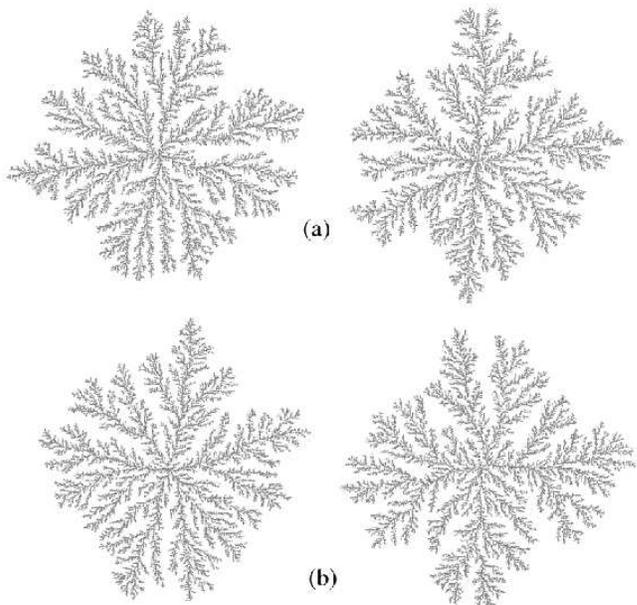}}
\end{center}
\vspace{-0.7cm}
\caption{Two clusters generated (a) using or (b) not the
optimization. In these simulations, lattices of size  $L=1000$ and
$\lambda=0.99$ were used.} \label{fig:otim}
\end{figure}

The first stage of the present work was to confirm the validity of the
previous defined algorithm. We simulate relatively small lattices containing
$10^3 \times 10^3$ sites with and without the optimization for $\lambda =
0.99$. In Fig. \ref{fig:otim}, comparisons between clusters generated with
(top) and without (bottom) the optimization are shown. Comparing the patterns,
one can see that they are statistically indistinguishable. Using the
mass-ratio method, the fractal dimensions of the patterns generated with and
without the algorithm were $d_f=1.70\pm 0.02$ and $d_f=1.72 \pm 0.02$,
respectively, and the exponents of the radius of gyration, defined by $r_g\sim
n^\zeta$ ($n$ is the number of cluster particles) \cite{Vicsekbook}, were
$\zeta=0.560\pm0.003$ and $\zeta=0.561\pm0.002$, respectively. These exponents
reinforce the algorithm validity. Concerning the computational time, a single
run to generate one of the clusters shown in Fig. \ref{fig:otim} without the
optimization takes about 1 hour in a 3 GHz Pentium IV, but the same simulation
is done in 10 minutes using the optimization. Therefore, even for small
lattices the simulation performance is greatly improved when our optimized
algorithm is used.

\begin{figure*}[hbt]
\begin{center}
\resizebox{15cm}{!}{\includegraphics{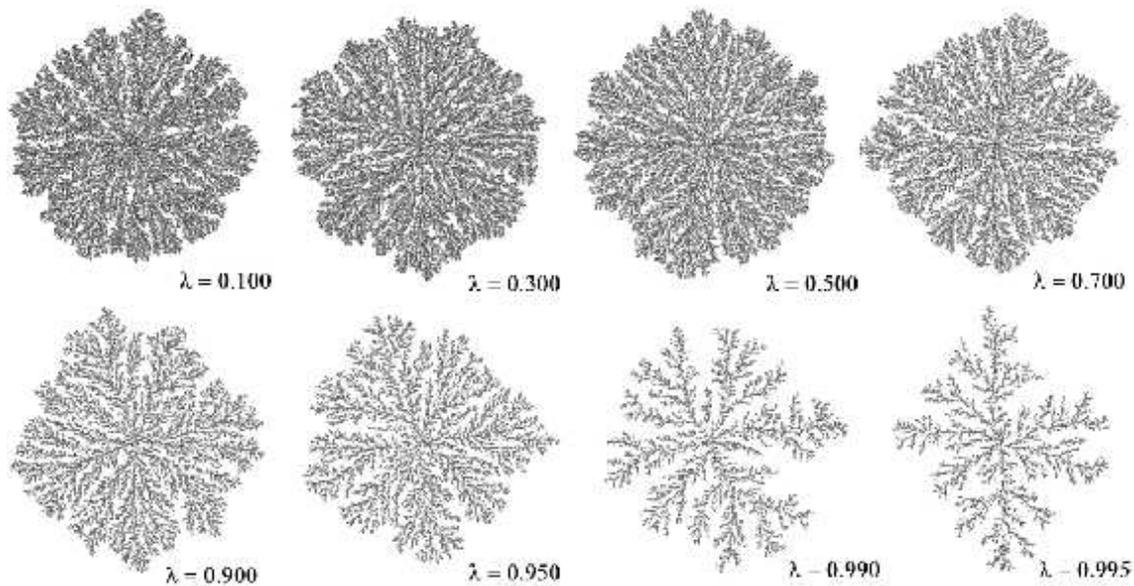}}
\end{center}
\vspace{-0.7cm}
\caption{Morphology transition between BA and DLA growth
patterns. The number of aggregated particles varies from $2.5\times10^5$
($\lambda=0.1$) to $5\times 10^4$ ($\lambda=0.995$).} \label{padroes}
\end{figure*}

Figure \ref{padroes} shows growth patterns for distinct $\lambda$ values.
These patterns were generated with the optimization at lattices containing
$10^3\times 10^3$ sites. The simulations stopped when the aggregate reaches
the lattice edge. A continuous transition from disordered and dense to
ramified clusters is observed. For small $\lambda$ values the patterns are
essentially BA-like but, the patterns become very similar to the DLA clusters
as $\lambda\rightarrow 1$. Indeed, the cluster generated with $\lambda=0.995$
is characterized by the square lattice anisotropy, a signature of the DLA
model \cite{Meakinbook, Vicsekbook}. However, one expects that all patterns
become asymptotically homogeneous with a finite characteristic size for the
empty regions.

In order to quantify the DLA to BA morphology transition, the mean particle density
in the inner regions of the cluster was evaluated. This mean density $\overline{\rho}(r)$ is
defined as the ratio between the number of occupied sites and the total number
of sites in a region delimited by a circle of radius $r$ centered at the
initial seed. Since one expects asymptotically nonfractal clusters, the
density must reach a finite value $\rho_0$ as $r\rightarrow \infty$.
Nevertheless, the approach to the constant density is very slow and takes a
scale invariant form
\begin{equation}
\overline{\rho}(r)=\rho_0+Ar^{-\gamma}.
\label{eq:rho}
\end{equation}
Here, $\gamma$ is a correction to the fractal dimension and $A$ a constant.
This scaling hypothesis was also used by Liang and Kadanoff \cite{Liang} to
study the driven ballistic aggregation, in which the particles trajectories
are in a single direction. They conclude that the $\gamma$ exponent is
nonuniversal, i.e., depends on the lattice structure.

\begin{figure}[hbt]
\begin{center}
\resizebox{8.5cm}{!}{\includegraphics{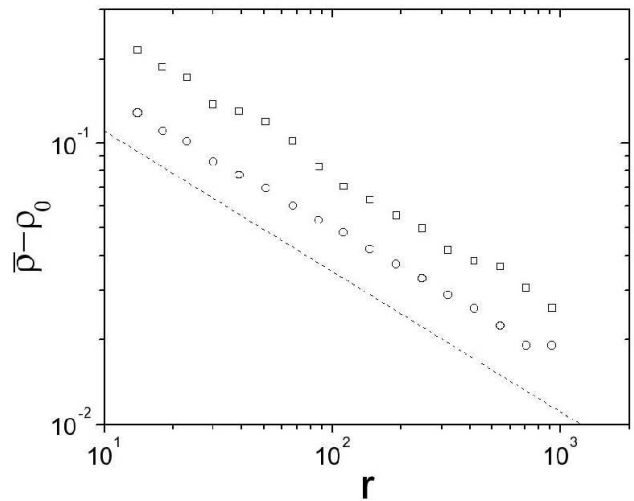}}
\end{center}
\vspace{-0.7cm}
\caption{Double-logarithm plots of $\overline{\rho}-\rho_0$ against $r$
for $\lambda=0.3$ (squares) and $\lambda=0.99$ (circles). The dashed line
corresponds to the slope $-0.46$. The linear fits of the data provide
$\gamma\approx 0.45$ for $\lambda=0.3$, and $\gamma\approx 0.47$ for
$\lambda=0.99$.} \label{fig:rho}
\end{figure}

\begin{figure}[hbt]
\begin{center}
\resizebox{8.5cm}{!}{\includegraphics{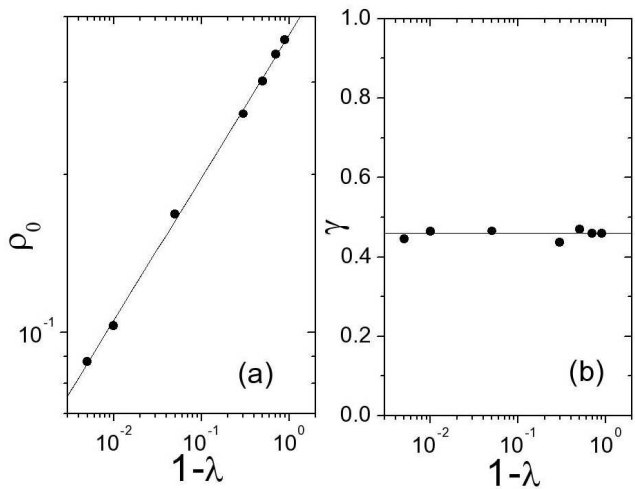}}
\end{center}
\vspace{-0.7cm}
\caption{ (a) Stationary density $\rho_0$ and (b) $\gamma$
exponent defined in Eq. (\ref{eq:rho}) as a function of the distance from the
transition point. The lattice size used was $L=5\times10^3$ and simulation
stopped when the cluster reaches the lattice edge.} \label{fig:beta}
\end{figure}

In Fig. \ref{fig:rho}, the double-logarithm plots of $\overline{\rho}-\rho_0$ as a
function of $r$ for distinct $\lambda$ values are shown. The density $\rho_0$
was obtained by searching for the best linear fit in the larger linear region.
To avoid the active region, we limited the fits to those data corresponding to
a half of the cluster sizes. Depending on the $\lambda$ value, lattices with
linear size $L=5\times10^3$ or $L=10^4$, and 10 to 20 independent runs were
used. One can observe a power law regime for $r>10$ showing that the approach
to the stationary value obeys Eq. \ref{eq:rho}. In Fig. \ref{fig:beta}, the
asymptotic density $\rho_0$ and the $\gamma$ exponent are shown as a function
of $1-\lambda$, the distance from the transition point. $\rho_0$ acts as an
order parameter, which vanishes at the critical point following a relation
$\rho_0\sim|1-\lambda|^\beta$. The exponent obtained from the data of Fig.
\ref{fig:beta}(a) was $\beta = 0.27(1)$, whereas the exponents obtained for
$L=2\times 10^3$ and $L=10^4$ were $\beta = 0.28(2)$ and $\beta =  0.26(1)$,
respectively. The numbers in parenthesis represent the uncertainties. In Fig.
\ref{fig:beta}(b), the $\gamma$ exponents for distinct $\lambda$ values are
shown. One can observe that $\gamma$ fluctuates around the value $0.46$. The
smaller value found was $\gamma\approx 0.43$ and the larger one $\gamma\approx
0.48$. Our simulations suggests that the $\gamma$ exponent is independent of
$\lambda$ and its value is $\gamma=0.46 \pm 0.02$. The error indicated in the
$\gamma$ value was evaluated through an average over the data of Fig.
\ref{fig:beta}(b).

In order to test the universality of the $\gamma$ and $\beta$ exponents we
studied two versions of the present model. In the first one, we used square
lattices, but the walkers stick to the cluster if they reach a nearest or a
next-nearest empty neighbor of an occupied site. In the second one, we used a
hexagonal lattice. Lattices with size $L=2000$ were used. The exponents for
the first modified version were $\beta=0.27\pm 0.02$ and $\gamma=0.49\pm0.03$,
and the exponents for hexagonal lattice were $\beta=0.28\pm0.02$ and
$\gamma=0.49\pm0.03$. These results lead us to conclude that these exponents
are universal.

\begin{figure}[hbt]
\begin{center}
\resizebox{8.5cm}{!}{\includegraphics{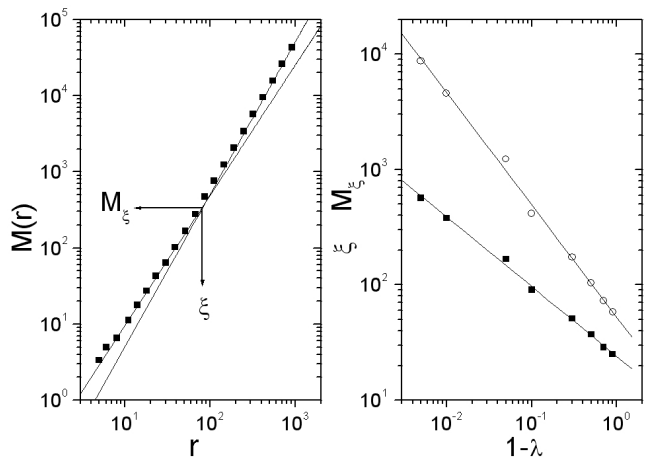}}
\end{center}
\vspace{-0.7cm}
\caption{ (a) Determination of the crossover between DLA- and
BA-like scaling regimes for $\lambda=0.90$. The straight lines represent the
slopes $1.71$ and $2$, i.e., the fractal dimensions for DLA and BA models,
respectively. (b) The crossover length (squares)  and the correspondent mass (circles)  as a function of the distance from the transition point. These results were obtained for $L=5\times10^3$.}
\label{fig:massa}
\end{figure}

The morphological transition between BA and DLA models was also characterized
by the crossover radius $\xi$ defined in Eqs. (\ref{eqM}) and (\ref{eqf}). The
number of particles $M(r)$ inside a region delimited by a circle of radius $r$
centered at origin was evaluated. The $M~vs.~r$ curves exhibit tenuous
crossovers determining the transition between DLA- and BA-like scaling
regimes. In Fig. \ref{fig:massa}(a), an example of this crossover is shown.
Since the growth patterns scale as DLA (BA) for small (large) length scales,
in order to evaluate the crossover $\xi$ we fitted the curves by power laws
$M(r)\sim r^{d_f}$, where $d_f=1.71$ and $d_f=2$ were used for the initial and
the final curve regions, respectively. The crossover lengths obtained through
this method are drawn as a function of the distance from the transition point
in Fig. \ref{fig:massa}(b). The length $\xi$ diverges at $\lambda=1$ following
a power law $\xi\sim |1-\lambda|^{-\nu}$, where $\nu=0.61(1)$. Moreover, the mass at the critical point diverges as $M_\xi\sim |1-\lambda|^{-\alpha}$, where $\alpha=0.97(2)$.

As discussed in section \ref{intro}, these crossovers between fractal and homogeneous patterns
occur due the crossover in the particle trajectories. However, the crossover
length of the walker trajectories is given by (see appendix)
\begin{equation}
\xi_w=\frac{\pi\lambda}{\sin(\pi\lambda)}-\frac{\sin(\pi\lambda)}{\pi\lambda},
\label{eq:xi}
\end{equation}
which diverges as $\xi_w\sim|1-\lambda|^{-1}$ for $\lambda\lesssim1$. Thus,
although the transition between DLA and BA models is due to the transition in
the walk dimensionality, the respective crossover lengths are not
proportional. 

Notice that Eq. (\ref{eq:rho}) describes the mean density behavior when $r \gtrapprox  \xi$. Moreover, $\overline{\rho}\sim r^{d_f-2}$ when $r\ll\xi$ due to the cluster fractality in this length scale. Thus, using Eq. (\ref{eq:rho}), we found that the mean density at the crossover can be written as
\begin{equation}
\overline{\rho}_\xi=A_1|1-\lambda|^\beta+A_2|1-\lambda|^{\nu\gamma},
\label{eq:rhoxi}
\end{equation}
where $A_1$ and $A_2$ are constants. But, the mean density at the crossover is given by
\begin{equation}
\overline{\rho}_\xi\sim \frac{M_\xi}{\xi^2}\sim|1-\lambda|^{2\nu-\alpha}.
\label{eq:rhoxi2}
\end{equation}
Comparing Eqs. (\ref{eq:rhoxi}) and (\ref{eq:rhoxi2}), we have that they are consistent only if
\begin{equation} 
\beta=\nu\gamma{\rm~and~} \beta=2\nu-\alpha.
\label{eq:scal}
\end{equation}
In agreement with the scaling relation (\ref{eq:scal}), the number of independent exponents are reduced from 4 to 2. Using the exponents measured for systems with size $L=5000$, we found $\nu\gamma=0.28(2)$ and $2\nu-\alpha=0.25(3)$, beside $\beta=0.27(1)$. The difference between these values is inside of the error margins indicated in the parenthesis. The large uncertainities obtained in the exponents (5\%-10\%) are originated in the difficulty in the determination of the exact crossover points.

\section{Conclusions}
\label{conclusions}

In the present work, we studied the transition between diffusion-limited
aggregation (DLA) and ballistic aggregation (BA) models. We used a model in
which the random walks in the DLA model are replaced by biased random walks
with a drift in a random direction. The drift is controlled by a parameter
$\lambda\in[0,1]$ that leads the model from BA ($\lambda=0$) to DLA
($\lambda=1$) (see Eq. (\ref{bias_walk})). Also, an efficient algorithm, which
allows large scaling analysis of the growth patterns, was introduced.

For any bias, the clusters are fractal (DLA-like) on the short length scales
whereas nonfractal patterns are obtained on the large ones. The transition
between DLA- and BA-like scaling regimes is determined by a characteristic
length $\xi$ that diverges as $\lambda\rightarrow 1$ following a power law
$\xi\sim |1-\lambda|^{-\nu}$, where $\nu= 0.61(1)$,(r) while the cluster mass
at the crossover follows the relation $M_\xi\sim |1-\lambda|^{-\alpha}$,
where $\alpha=0.97(2)$. This crossover was not
numerically determined in similar previous works. The density in the inner
regions of the cluster reaches an asymptotic value
$\rho_0\sim|1-\lambda|^\beta$, where $\beta = 0.26(1)$. However, this approach
is slow and follows a power law decay with a universal exponent
$\gamma=0.46(2)$ independent on the drift. These exponents obey the scaling relations $\beta=\nu\gamma {\rm~and~} \beta=2\nu-\alpha$.

It is worth to stress two main contributions of the present work. The first
one is the development of an algorithm that can be used to study other models
with biased random walks as for example those related to deposition processes
\cite{CastroPRE,Kim}, for which the determination of universality classes is a
hard work. The second one is the careful quantitative characterization of the
transition between DLA and BA growth models that, in our knowledge, was not
previously done. The understanding of these crossovers can be an essential
tool in the analysis of real fractals, which always exhibit scaling on limited
ranges.

\begin{acknowledgments} We would like to thank M. L. Martins and A. P. F. Atman
for the critical reading of the manuscript. This work was partially supported by
the CNPq Brazilian agency.\end{acknowledgments}

\appendix*
\section{Demonstration of Eq. (\ref{eq:xi}) }
For sake of simplicity, we consider Eq. (\ref{bias_walk}) with a drift
direction $\varphi=0$ and $x_0=y_0=0$. Iterating Eq. (\ref{bias_walk}) for $n$
steps, we found
\begin{equation}
\langle x_n \rangle = \sum_{i=1}^{n} \langle \cos(\lambda\theta_i) \rangle,
\label{eq:xm}
\end{equation}
and
\begin{equation}
\langle x_n^2 \rangle=\sum_{i=1}^{n}\sum_{j=1}^{n}\langle\cos(\lambda\theta_i)\cos(\lambda\theta_j) \rangle.
\label{eq:xm2}
\end{equation}
But,
\begin{equation}
\langle \cos(\lambda\theta_i) \rangle = \frac{1}{2\pi}\int_{-\pi}^{\pi}\cos(\lambda\theta) d\theta = \frac{\sin(\pi\lambda)}{\pi\lambda}.
\label{eq:cos1}
\end{equation}
and
\begin{eqnarray}
\langle \cos(\lambda\theta_i)\cos(\lambda\theta_j) \rangle =
\left[\frac{\sin(\pi\lambda)}
{\pi\lambda}\right]^2(1-\delta_{ij})\nonumber\\+\frac{1}{2}\left[\frac{\sin(2\pi\lambda)}
{2\pi\lambda}+1\right]\delta_{ij}, \label{eq:cos2}
\end{eqnarray}
where $\delta_{ij}$ is the Kronecker delta.

Substituting Eqs. (\ref{eq:cos1}) and (\ref{eq:cos2}) in Eqs. (\ref{eq:xm}) and (\ref{eq:xm2}), respectively, we found
\begin{equation}
\langle x_n \rangle = n \frac{\sin(\pi\lambda)}{\pi\lambda}
\end{equation}
and
\begin{equation}
\sigma_x^2(n)=n\left \{\frac{1}{2} \left[1+\frac{\sin(2\pi\lambda)}
{2\pi\lambda}\right]- \left[\frac{\sin(\pi\lambda)} {\pi\lambda}\right]^2
\right \},
\end{equation}
where $\sigma_x^2(n)=\langle x_n^2\rangle-\langle x_n\rangle^2$ is the
variance of the coordinate $x_n$. With a similar analysis, we obtained
\begin{equation}
\langle y_n \rangle =0
\end{equation}
and
\begin{equation}
\sigma_y^2(n)=n\frac{1}{2} \left[1-\frac{\sin(2\pi\lambda)}{2\pi\lambda}\right].
\end{equation}
Thus, the  walk mean displacement $r(n)=\sqrt{\langle x_n\rangle^2 + \langle
y_n\rangle^2} $ and the variance $\sigma^2(n)=\sigma_x^2+\sigma_y^2$ are given
by
\begin{equation}
r(n)=n \frac{\sin(\pi\lambda)}{\pi\lambda}
\label{eq:rn2}
\end{equation}
and
\begin{equation}
\sigma(n)=n^{1/2}\left \{1-\left[\frac{\sin(\pi\lambda)} {\pi\lambda}\right]^2 \right \}^{1/2}.
\label{eq:sigman2}
\end{equation}

\begin{figure}[hbt]
\begin{center}
\resizebox{8.5cm}{!}{\includegraphics{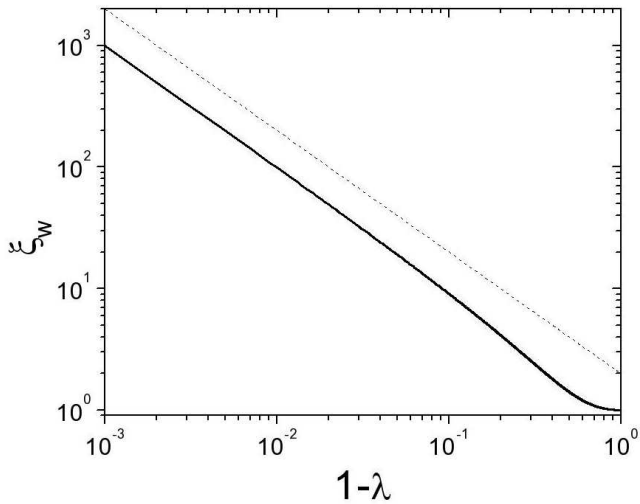}}
\end{center}
\vspace{-0.7cm}
\caption{Double-logarithm plot of the crossover length $\xi_w$
as a function of $1-\lambda$. The solid line represents the data obtained from
Eq. (\ref{eq:xi2}) and the dashed line has a slope $-1$.} \label{fig:cross}
\end{figure}

The crossover of the walk dimensionality ($d=2$ for short times and $d=1$ for
long ones) occurs when $r(n) \sim \sigma(n)$. Making equal Eqs. (\ref{eq:rn2})
and (\ref{eq:sigman2}), we obtained an estimative of the characteristic number
of steps $\mathcal{N}$ necessary for the crossover
\begin{equation}
\mathcal{N}=\left[\frac{\pi\lambda}{\sin(\pi\lambda)}\right]^2-1.
\end{equation}
Thus, the characteristic crossover length is
\begin{equation}
\xi_w=r(\mathcal{N})=\frac{\pi\lambda}{\sin(\pi\lambda)}-\frac{\sin(\pi\lambda)}{\pi\lambda}.
\label{eq:xi2}
\end{equation}
In Fig. (\ref{fig:cross}), $\xi_w$ is plotted as a function of $1-\lambda$.
Expanding Eq. (\ref{eq:xi2}) around $\lambda=1$, we found $\xi_w\sim
|1-\lambda|^{-1}$.

\end{document}